\documentstyle[times,graphics,astrobib,amssymb]{mn2e}

\newcommand{\be}{\begin{equation}}
\newcommand{\e}{\end{equation}}
\newcommand{\bear}{\begin{eqnarray}}
\newcommand{\ear}{\end{eqnarray}}

\newcommand{\f}{\frac}
\newcommand{\de}{{\rm d}}

\begin{document}

\title[Reionization sources]
{Searching for the reionization sources}
\author[Choudhury \& Ferrara]
{T. Roy Choudhury$^{1}$\thanks{E-mail: chou@ast.cam.ac.uk}~
and
A. Ferrara$^{2}$\thanks{E-mail: ferrara@sissa.it}\\
$^{1}$Institute of Astronomy, Madingley Road, Cambridge CB3 0HA, UK\\
$^{2}$SISSA/ISAS, via Beirut 2-4, 34014 Trieste, Italy}

\maketitle

\date{\today}

\begin{abstract}
Using a reionization model simultaneously accounting for a number of experimental data sets, we investigate the nature and properties 
of reionization sources.  Such model predicts that hydrogen reionization starts at $z \approx 15$, is initially driven by metal-free
(PopIII) stars, and is 90\% complete by $z \approx 8$.  We find that a fraction $f_\gamma >80$\% of the ionizing 
power at $z \ge 7$ comes from haloes of mass $M< 10^9 M_{\odot}$ predominantly harboring PopIII stars; a 
turnover to a PopII-dominated phase occurs shortly after, with this population, residing in $M>10^9 M_\odot$ haloes, 
yielding $f_\gamma\approx 60$\% at $z=6$. Using Lyman-break broadband dropout techniques, $J$-band detection of sources 
contributing to 50\% (90\%) of the ionizing power at $z \sim 7.5$ requires to reach a magnitude 
$J_{110,{\rm AB}} = 31.2 (31.7)$, where $\sim 15 (30)$ (PopIII) sources/arcmin$^2$ are predicted. We conclude
that $z>7$ sources tentatively identified in broadband surveys are relatively massive ($M\approx 10^9 M_\odot$) and rare objects 
which are only marginally ($\approx 1$\%) adding to the reionization photon budget. 
\end{abstract}
\section{Introduction}

The study of reionization received a big boost due to availability of a variety of observational data accumulated over the past few years
(for reviews see \citeNP{fck06} and \citeNP{cf06a}); additional progresses are soon expected from a number of different ground-based (LOFAR, MWA, 
PAST, SKA, ALMA) and space-born (JWST, PLANCK, GLAST) experiments.  

The available results have allowed us to build self-consistent reionization scenarios that are able to account simultaneously
for a number of observables (redshift evolution of Lyman-limit absorption systems, GP and electron scattering optical depths, 
mean temperature of the intergalactic medium (IGM),  and cosmic star formation history;
\nocite{cf05,cf06b} Choudhury \& Ferrara 2005, 2006; hereafter CF05 and CF06 respectively). To summarize the emerging picture,
the most favorable model is one in which hydrogen reionization was an extended process starting around $z \approx 15$ and 
being 90\% complete by $z \approx 8$.  This (early) reionization model was also shown not to be in conflict with the Gunn-Peterson
optical depth evolution deduced from QSO absorption line experiments at $z \gtrsim 6$ both by using the statistics of dark gaps in 
the Ly$\alpha$ transmitted flux \cite{gcf06} and through radiative transfer simulations of ionized regions around QSOs \cite{mgfc07}. 
According to the model, reionization is initially driven by metal-free stars in low mass ($M < 10^8 M_{\odot}$) haloes; the 
conditions for the formation of these objects are soon erased by the combined action of chemical and radiative feedback at $z < 10$.
As a consequence, the photoionizing power (and therefore integrated luminosity) of these sources is most significant around $z \approx 8-12$.  

In spite of this successful overall picture, relatively less attention has been devoted so far to the observable properties of the sources 
responsible for cosmic reionization. Our main aim in this work is to fill such gap by providing quantitative guidelines for observers searching 
for the first cosmic light sources.   Specifically, we use the CF05 and CF06 model to estimate the IR fluxes and magnitude-limited counts
of the primary reionization sources.  

\begin{figure*}
\rotatebox{270}{\resizebox{0.4\textwidth}{!}{\includegraphics{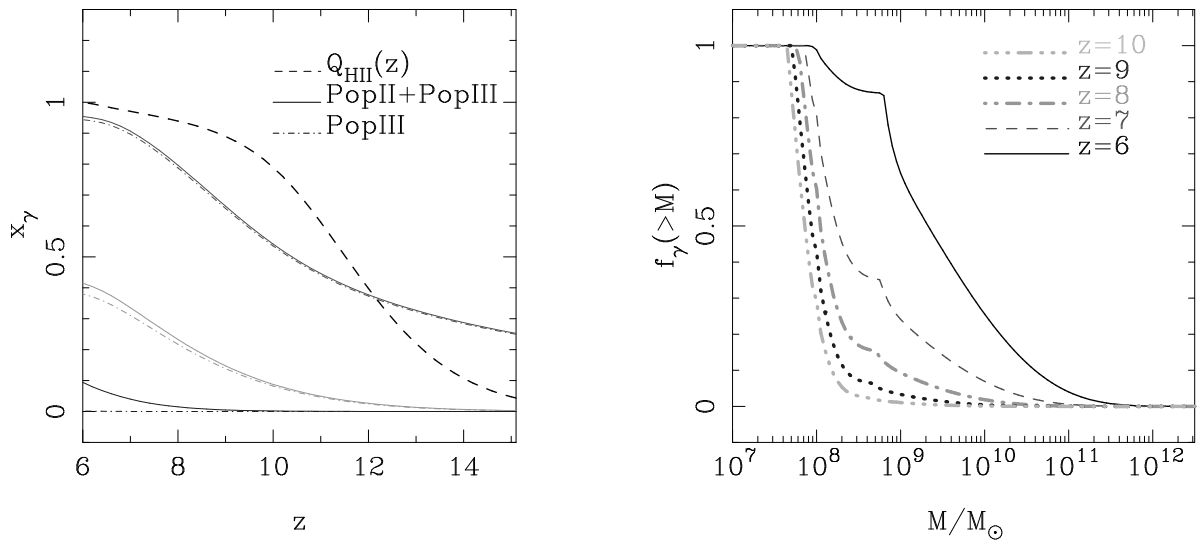}}}
\caption{{\it Left}: Number of ionizing photons per H-atom contributed by haloes of different mass 
in a fraction of the Hubble time $t_H(z)$ equal to the recombination time $t_{\rm rec}(z)$ as a function of 
redshift $z$ for different halo masses and for different stellar populations. The solid and dot-dashed pair of 
curves, from top to bottom, represent haloes  of masses $10^7 M_{\odot} < M < 10^8 M_{\odot}$, $10^8 M_{\odot} < M < 10^9 M_{\odot}$, 
and $M >  10^9 M_{\odot}$ respectively. The dashed line represent the evolution of the volume filling factor $Q_{\rm HII}$ of ionized regions.
{\it Right}: Cumulative fraction of the ionizing power $f_{\gamma}$ contributed by haloes of mass $>M$. The curves from 
right to left correspond to $z = 6,7,8,9,10$ respectively.
}
\label{fig:halofrac}
\end{figure*}
\vspace{-0.5cm}
\section{Basic features of the model}
The main features of the semi-analytical model used in this work\footnote{
Throughout the paper, we use the best-fit cosmological parameters from the 3-year WMAP data \cite{sbd++06}, i.e., 
a flat universe with $\Omega_m = 0.24$, $\Omega_{\Lambda} = 0.76$, and $\Omega_b h^2 = 0.022$, and $h=0.73$.  The 
parameters defining the linear dark matter power spectrum are $\sigma_8=0.74$, $n_s=0.95$, $\de n_s/\de \ln k =0$.}
could be summarized along the following points (for a more detailed description
see CF05 and CF06). The model accounts for IGM inhomogeneities by adopting a lognormal distribution according to the method outlined in \citeN{mhr00}; 
reionization is said to be complete once all the low-density regions (say, with overdensities $\Delta < \Delta_{\rm crit} \sim 60$) are ionized. 
The mean free path of photons is thus determined essentially by the distribution of high density regions.  We follow the ionization and thermal histories 
of neutral, HII and HeIII regions simultaneously and self-consistently, treating the IGM as a multi-phase medium. 

Three types of reionization sources have been assumed: (i) metal-free (i.e. PopIII) stars having a Salpeter IMF in the mass 
range $1 - 100 M_{\odot}$: they dominate the photoionization rate at high redshifts; (ii) PopII stars with sub-solar metallicities 
also having a Salpeter IMF in the mass range $1 - 100 M_{\odot}$; (iii) QSOs, which are 
significant sources of hard photons at $z \lesssim 6$; they have negligible effects on the IGM at higher redshifts. Note that there 
is no compelling reason to rule out the possibility that the IMF of PopIII stars was top-heavy; however, as this is not required by 
current data, we limit ourselves to the most conservative model.

Reionization by UV sources is accompanied by photo-heating of the gas, which can result in a suppression of star formation in low-mass 
haloes. We compute such (radiative) feedback self-consistently from the evolution of the thermal properties of the IGM. Furthermore, 
the chemical feedback inducing the PopIII $\rightarrow$ PopII transition is implemented according to the detailed study by Schneider 
et al. (2006) in which a merger-tree ``genetic'' approach is used to determine the termination of PopIII star formation in a metal-enriched halo.

The predictions of the model are compared with a wide range of observational data sets, namely, (i) redshift evolution of Lyman-limit absorption 
systems \cite{smih94}, (ii)  IGM Ly$\alpha$ and Ly$\beta$ optical depths \cite{songaila04}, (iii) electron scattering optical depth \cite{sbd++06}, 
(iv) temperature of the mean intergalactic gas \cite{stle99}, (v)  cosmic star formation history \cite{nchos04}, and (vi) source number counts at $z 
\approx 10$ from NICMOS HUDF \cite{bitf05}. 

The best-fit reionization model is characterized by a PopII (PopIII) star-forming efficiency $\epsilon_{*,{\rm II}} = 0.1 (\epsilon_{*,{\rm III}} = 0.03)$
and escape fraction  $f_{\rm esc, II} = 0.01 (f_{\rm esc, III} = 0.68)$ (keeping in mind that $f_{\rm esc, II}$ and $f_{\rm esc, III}$ 
are not independent).\footnote{The values are slightly different from those quoted in CF06 because of an improved
likelihood analysis. The qualitative results and main conclusions are unaffected by this modification.} The resultant value of the electron scattering 
optical depth is $\tau_{\rm el} = 0.1$

The data constrain the reionization scenario quite tightly. We find that hydrogen reionization starts at $z \approx 15$ driven by metal-free 
(PopIII) stars, and it is 90 per cent complete by $z \approx 8$.  This can be seen from the dashed curve in Fig.   \ref{fig:halofrac} which
represents the evolution of the volume filling factor $Q_{\rm HII}(z)$ of ionized regions.  After a rapid initial phase, the growth of the
volume filled by ionized regions slows down at $z \lesssim 10$ due to the combined action of chemical and radiative feedback,  
making reionization a considerably extended process completing only at $z \approx 6$.
\vspace{-0.5cm}
\section{Properties of the reionization sources}
\begin{figure*}
\rotatebox{270}{\resizebox{0.45\textwidth}{!}{\includegraphics{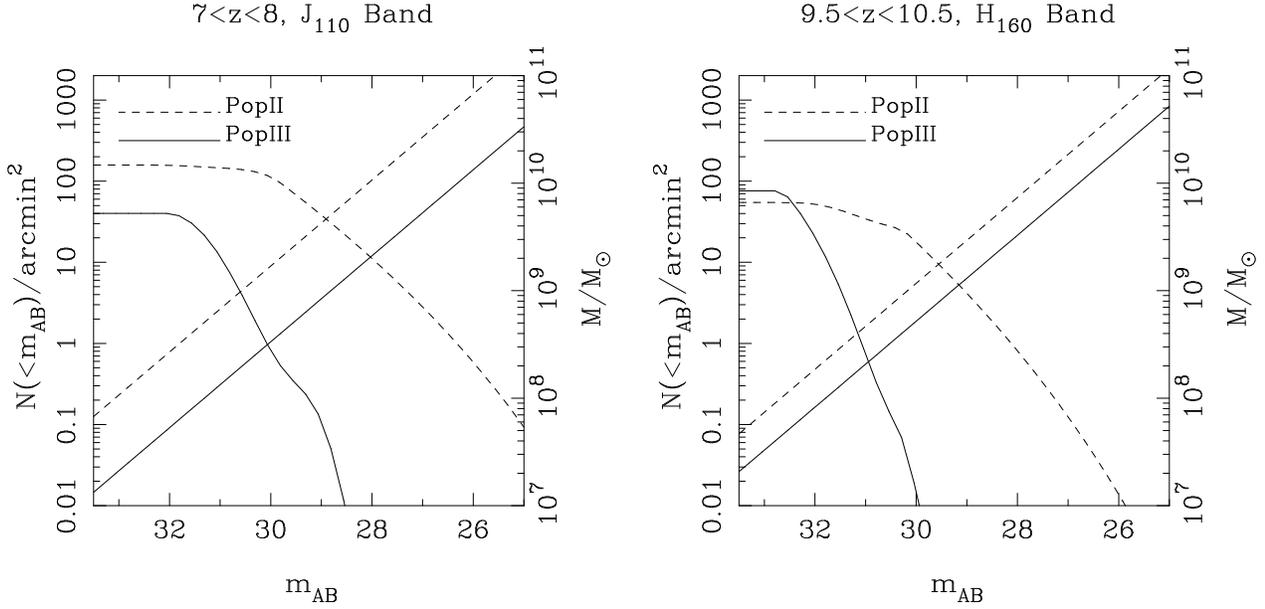}}}
\caption{Number density of reionization sources (PopII and PopIII) as a function of the limiting magnitude, $m_{\rm AB}$,
at $7 < z < 8$ observed in $J_{110}$ band (left panel) and at $9.5 < z < 10.5$ observed in $H_{160}$ band (right).
Halo masses corresponding to $m_{\rm AB}$ for the particular redshift under consideration [given by equations (\ref{eq:f_nu_0}) 
and (\ref{eq:m_ab})] are shown by the straight lines and the relevant values can be read off from the right vertical axis.
}
\label{fig:sourcefrac}
\end{figure*}
Having identified the most probable reionization scenario, we can now confidently 
determine the properties of the reionization sources governing it. Let us start by 
defining the quantity
\be
x_{\gamma}(z) \equiv \f{n_{\gamma}(z)}{n_H} \f{t_{\rm rec}(z)}{t_H(z)}
\e
as the number of ionizing photons per H-atom contributed by haloes in the mass range $[M_{\rm min}, M_{\rm max}]$
in a fraction of the Hubble time $t_H(z)$ equal to the recombination time $t_{\rm rec}(z)$.
By construction, the IGM is reionized when $x_{\gamma} \gtrsim 1$.  
The quantity $n_H$ is the comoving number density of hydrogen 
atoms while $n_{\gamma}$
is the time-integrated comoving photon density, calculated using the relation
\be
n_{\gamma}(z) = \int_0^{t(z)} \de t ~ \dot{n}_{\gamma}(M_{\rm min} : M_{\rm max}, t)
\e
where $\dot{n}_{\gamma}(M_{\rm min} : M_{\rm max}, t)$ is the ionizing photon comoving emissivity 
from haloes within $[M_{\rm min}, M_{\rm max}]$.

The plot of $x_{\gamma}(z)$ for different mass ranges is shown in Fig.   \ref{fig:halofrac}.  The lower mass 
$10^7$-$10^8 M_{\odot}$ haloes  dominate the photon production rate at early redshifts providing about 0.25 photon/H-atom on
the fractional recombination timescale. These objects produce the first ionized regions, are preferentially metal-free,
and therefore mostly harbor PopIII stars of high specific ionizing power. 
At $z < 8$, the contribution from PopIII haloes decreases because their formation is hampered by the
heating associated with radiative feedback. As a result, the progress of  ionization fronts relies on photons emitted by 
more massive haloes with $ M> 10^9 M_{\odot}$. It can be seen from Fig.1 that such high mass haloes do not host PopIII stars 
as they form from the merging of already polluted progenitors, a result of the ``genetic'' transmission of chemical feedback 
(Schneider et al. 2006).  This combination of radiative and chemical feedback makes the reionization process quite extended and its 
completion has to wait until $z \approx 6$ when the PopII stars (and QSOs, not shown in the Figure; see CF06 for details) dominate  
the photoionization rate. 

Additional insights on the source properties may be gained by considering the fractional instantaneous contribution of 
haloes above a certain mass, 
\be
f_{\gamma}(>M, z) \equiv \f{\dot{n}_{\gamma}(>M, z)}{\dot{n}_{\gamma}(z)},
\e
also shown for different redshifts in the right hand panel of Figure \ref{fig:halofrac}.  
The results shown emphasize again that $>80$\% of the ionizing power at $z \ge 7$ is provided by haloes with masses $< 10^9 M_{\odot}$ 
which are predominantly harboring PopIII stars. A turnover to a PopII-dominated reionization phase occurs shortly after, with 
this population, residing in $M>10^9 M_\odot$ haloes, producing $\approx 60$\% of the ionizing photons at $z=6$. In conclusion, 
PopIII stars and small galaxies initiate reionization at high redshift and remain important until they are overcome by PopII stars 
and QSOs below $z=7$. 

\vspace{-0.5cm}
\section{Source counts at high redshift}

Having determined which haloes contribute most significantly to the ionizing power at a particular redshift,
we now address the detectability of these sources in broadband imaging surveys and the optimal strategies to do so. A halo of (dark matter) mass $M$ at a redshift
$z$ will emit a flux given by \cite{sf06}
\be
F_{\nu_0} = \f{\epsilon_{*} (\Omega_b/\Omega_m) M}{4 \pi d_L^2(z') \Delta \nu_0} 
\int \de \nu' ~ l_{\nu'}(\Delta t)~ {\rm e}^{-\tau_{\rm eff}(\nu_0,z=0,z')}
\label{eq:f_nu_0}
\e
where $\epsilon_{*}$ is the star-forming efficiency of the population under consideration, $l_{\nu'}(\Delta t)$ is a template
specific luminosity for the stellar population of age $\Delta t=t_{z'}-t_{z''}$ (the time elapsed between the two redshifts), 
$d_L(z')$ is the luminosity distance and $\Delta \nu_0$ is the instrumental bandwidth. The quantity $\tau_{\rm eff}(\nu_0,z=0,z')$ is 
the effective optical depth due to intervening gas at $\nu_0$ between $z'$ and $z = 0$. The main contribution to $\tau_{\rm eff}(\nu_0,z=0,z')$ comes
from the combined blanketing of the Lyman series lines in the emitter rest frame wavelength range 
$912{\rm \AA} <\lambda < 1216 {\rm \AA}$ and from the continuum absorption from neutral hydrogen in the range $\lambda <$ 912\AA.
Both the contributions to the opacity can be calculated self-consistently from the (modified) lognormal density distribution (CF05) assumed in our model.
For calculating the template luminosity $l_{\nu}$, we use stellar population models (having metallicity $Z = 0.004 = 0.2 Z_{\odot}$) of \citeN{bc03} for PopII stars and of \citeN{schaerer02} 
for PopIII stars. 
Note that the star-forming efficiency $\epsilon_{*}$ in the above equation
is fixed by constraining the reionization history and is 
{\it not} a free parameter as far as the calculation of the source counts in this work is concerned.
The flux $F_{\nu_0}$ can be transformed into a magnitude in the AB system:
\be
m_{\rm AB} = -2.5 \log_{10} \left(\f{F_{\nu_0}}{{\rm erg~s}^{-1} {\rm cm}^{-2} {\rm Hz}^{-1}}\right) - 48.6
\label{eq:m_ab}
\e

The number of sources within a redshift interval $[z_{\rm min}, z_{\rm max}]$ observed in a solid angle $\de \Omega$ having a flux larger than
$F_{\nu_0}$ is 
\be
N(>{F_{\nu_0}})
= \int_{z_{\rm min}}^{z_{\rm max}}
\de z' \f{\de V}{\de z' \de \Omega} 
\int_{F_{\nu_0}}^{\infty} \de F'_{\nu_0}
\f{\de n}{\de F'_{\nu_0}} (F'_{\nu_0}, z')
\e
where $\de V/\de z' \de \Omega$ denotes the comoving volume element per unit redshift per unit solid angle, and 
\be
\f{\de n}{\de F'_{\nu_0}} (F'_{\nu_0}, z') = \int_{z'}^{\infty}
\de z'' \f{\de M}{\de F'_{\nu_0}}(F'_{\nu_0}, \Delta t)
\f{\de^2 n}{\de M \de z''}(M,z'')
\e
is the comoving number of objects at redshift $z'$ with observed flux within $[F'_{\nu_0}, F'_{\nu_0} + \de F'_{\nu_0}]$.
The quantity $\de^2 n/\de M \de z''$ gives the formation rate of haloes  of mass $M$, which we obtain from our reionization model. 

For definiteness, let us consider two bands corresponding to the NICMOS observations of the HUDF \cite{bti++04,bitf05},
namely $J_{110}$ and $H_{160}$; these broadband filters are appropriate to detect Lyman-break dropout sources at $z \sim 7.5$ and $z \sim 10$, respectively. 
The main results of the calculation are shown in Fig. \ref{fig:sourcefrac} where we plot the number of sources (i.e. galaxies
powered by either PopII or PopIII stars) observed in a particular redshift interval and band as a function of the limiting magnitude $m_{\rm AB}$. 
The halo masses corresponding to $m_{\rm AB}$ for the particular redshift under consideration [given by equations (\ref{eq:f_nu_0}) and 
(\ref{eq:m_ab})] are shown by the straight lines and the relevant values can be read off from the right vertical axis.

From Fig. 2, we deduce that, at the currently achieved sensitivity limit of $\sim 28$ AB magnitude, our best-fit model predicts 
$\sim 10 (1)$ Lyman-break sources per arcmin$^2$ at $z \sim 7.5 (10)$ observable 
in the $J_{110}$ ($H_{160}$) band.  Note that all these sources are bright PopII star forming haloes having masses $\sim 10^{10} M_{\odot}$. 
In the previous Section we found that such sources provide only a negligible ($\approx 1\%$) contribution to reionization at $z > 6$.

Some of these sources have been tentatively identified in broadband observations. For example, observations of HUDF using the NICMOS filter 
\cite{bti++04,bitf05} at AB magnitude limit $\sim 28$ reveal $\sim 1$ source per arcmin$^2$ at $z \sim 7.5$ and $< 1$ source 
per arcmin$^2$ at  $z \sim 10$ respectively. On the other hand, deep near-IR photometry of two lensing clusters (A1835 and AC114) 
obtained with ISAAC/VLT \cite{rpslk06} seem to indicate $\sim 1$ source per arcmin$^2$ at $z \sim 7-10$ at a magnitude limit equivalent 
to $H_{160, {\rm AB}} \sim 26$, which is considerably higher than the NICMOS results.   Stringent constraints are difficult to obtain from theoretical models as there remains 
considerable confusion regarding the actual number of reliable detections.  However, although the actual source count is disputed, 
it is most likely that these tentatively detected sources are bright massive haloes which are not significant for reionization.

The sources responsible for the bulk of reionization are galaxies formed inside low mass $< 10^8 M_{\odot}$ haloes and powered by PopIII stars. 
By inspecting the right hand panel of Fig. \ref{fig:halofrac} we conclude that in order to observe sources which contribute to 50\% (90\%) 
of the ionizing power at $z \sim 7.5$, it is necessary to observe PopIII sources of mass $<10^{8.1} M_{\odot} (<10^{7.9} M_{\odot})$. This
in turn requires to reach a magnitude sensitivity of $J_{110,{\rm AB}} = 31.2 (31.7)$ [left panel of Fig. \ref{fig:sourcefrac}], currently
beyond the reach of any instrument.  Interestingly, one can expect to observe $\sim 15 (30)$ PopIII sources per arcmin$^2$ once such sensitivities 
are reached.  Similar conclusions can be drawn from the right hand panel for sources at $z \sim 10$ observable in the $H_{160}$ band. 
For detecting sources which contribute to 50\% (90\%) of the ionizing power, the magnitude sensitivity required would be 
$H_{160,{\rm AB}} = 32.2 (32.5)$, which would again correspond to halo masses of $10^{7.8} M_{\odot} (10^{7.7} M_{\odot})$. Hence, the detection 
of reionization sources would require sensitivities better than 32 AB magnitudes in this band.

\vspace{-0.5cm}
\subsection{Variants of the best-fit model}
\begin{figure}
\rotatebox{270}{\resizebox{0.45\textwidth}{!}{\includegraphics{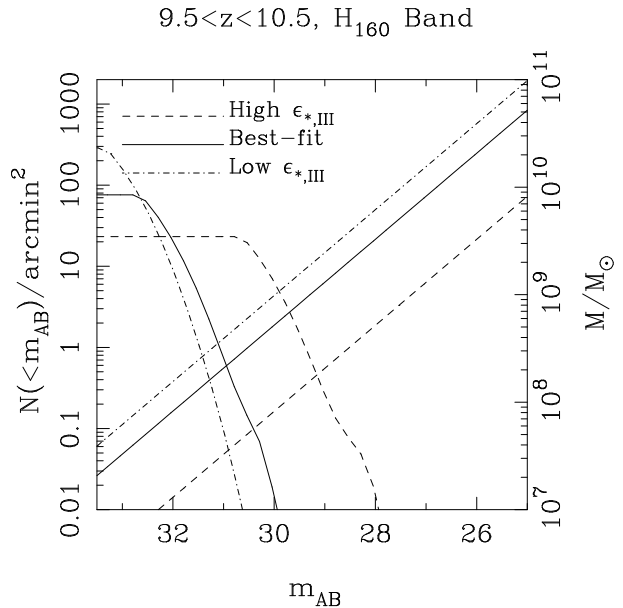}}}
\caption{Number density of PopIII sources as a function of the limiting magnitude, $m_{\rm AB}$,
at $9.5 < z < 10.5$ observed in $H_{160}$ band. The plots are for three models
of PopIII star-formation as discussed in the text.
Halo masses corresponding to $m_{\rm AB}$ for the particular redshift under consideration [given by equations (\ref{eq:f_nu_0}) 
and (\ref{eq:m_ab})] are shown by the straight lines and the relevant values can be read off from the right vertical axis.
}
\label{fig:sourcefrac2}
\end{figure}

In this subsection, we try to obtain some indication about how much can the source count estimates vary without violating 
any other observational constraints. We first note that the star-forming efficiency of PopII sources are 
quite stringently constrained by low-redshift observations, e.g., the cosmic SFR and the QSO absorption lines; hence their 
numbers cannot vary significantly. The situation is markedly different for PopIII stars as there are very few observations 
constraining their properties. In order to calculate the variation in PopIII source counts, we consider two extreme limits 
of the parameter $\epsilon_{*,{\rm III}}$ which are essentially determined by the bounds on $\tau_{\rm el}$:

(i) The high-$\epsilon_{*,{\rm III}}$ model: This model is characterized
by the parameters $\epsilon_{*,{\rm II}} = 0.1, \epsilon_{*,{\rm III}} = 0.15, f_{\rm esc, II} = 0.0,  f_{\rm esc, III} = 0.4$. 
The resultant $\tau_{\rm el}$ is 0.12 (the 1-$\sigma$ upper limit given by WMAP3; \citeNP{sbd++06}).
The initial stages of reionization in this case proceeds much faster
than in the best-fit model and 90 per cent of the IGM is
ionized by $z \approx 10.5$. However, because of the high efficiency
of the PopIII stars, the radiative feedback is more severe and hence the
reionization gets extended till $z \approx 6$
as in the best-fit model.

(ii) The low-$\epsilon_{*,{\rm III}}$ model: This model is characterized
by the parameters $\epsilon_{*,{\rm II}} = 0.1, \epsilon_{*,{\rm III}} = 0.01, f_{\rm esc, II} = 0.04,  f_{\rm esc, III} = 0.06$. 
The resultant $\tau_{\rm el}$ is 0.06 (the 1-$\sigma$ lower limit given by WMAP3; \citeNP{sbd++06}).
This model corresponds to the minimum ionizing power contribution 
required from PopIII stars. This model has an important qualitative difference from the high-$\epsilon_{*,{\rm III}}$ model; 
because of such low PopIII efficiency, almost
all of the reionization process is driven by the PopII stars. In fact, the 
lack of ionizing power at high-$z$ makes reionization start much later. Also, 
the effects of feedback are minor in this model; as a result, the reionization
occurs rapidly and shortly before $z=6$.

We now proceed to discuss how these models differ in their predictions for PopIII source counts (Fig. \ref{fig:sourcefrac2}). 
Since the effects of PopIII stars are most noticeable at $z \gtrsim 10$, we focus on the $H_{160}$ band.
The first important point is that the number of bright sources increases with $\epsilon_{*,{\rm III}}$,  a
direct consequence of the higher specific SFR within dark matter haloes. This would imply that one can observe 
most of the reionization sources at a magnitude limit of $H_{160, {\rm AB}} \sim 30.5$, a limit almost certainly
attained by near future experiments. However, the number of sources detected at fainter magnitudes is smaller for
the high-$\epsilon_{*,{\rm III}}$ model than for the other two (best-fit and high-$\epsilon_{*,{\rm III}}$) models. This is because 
the efficient star formation at high-$z$ makes the radiative feedback more effective, hence effectively quenching star 
formation in low-mass haloes and suppressing the number of faint sources.
Of course, one finds a rise in the number of faint sources for the
low-$\epsilon_{*,{\rm III}}$ model when the feedback has the least effect; 
one should be able to observe, in principle, $\gtrsim 100$ sources at 
$H_{160, {\rm AB}} \sim 33$. However, one should keep in mind that 
for low values of 
$\epsilon_{*,{\rm III}} \approx 0.01$, 
PopIII stars are {\it not} the main driving force for reionization anyway, and the main focus should be the source counts for PopII sources. In fact, 
we find that sources contributing to 90\% of the ionizing power at 
$z \approx 7.5$ would be the 
PopII stars with haloes of mass $< 10^8 M_{\odot}$; they 
can be observed with a sensitivity limit of 
$J_{110,{\rm AB}} \sim 33$, which is quite difficult with 
near-future experiments.

It is thus clear from the above discussion that the luminosity
function of PopIII sources at $z \approx 10$ could be important
for constraining the star-forming efficiency of PopIII stars and also 
for some indirect understanding of feedback effects.

\vspace{-0.5cm}
\section{Discussion}

We have used the self-consistent model of CF06, which is consistent
with a variety of observations, to estimate the number 
of sources (galaxies) at $z \sim 7-10$. According to our analysis:
\begin{itemize}

\item The best-fit model predicts $\gtrsim 1$ sources at $z \sim 7-10$
per arcmin$^2$ at a sensitivity limit of $\sim 28$ AB magnitude, roughly
consistent with present observations. However, these sources are 
bright massive ($\gtrsim 10^{10} M_{\odot}$) haloes  forming
PopII stars which have negligible contribution to reionization
of the IGM.

\item Reionization at $z > 6$ is actually driven by PopIII
stars in low mass ($< 10^{8} M_{\odot}$) haloes. The required
sensitivity to detect these sources would be 31-32 AB magnitude.

\item In case the star-forming efficiency of PopIII stars
is higher than what is assumed in our best-fit model, the reionization
sources can be detected with sensitivities of 30 AB magnitude; however, 
the number of sources detected could be smaller than the
best-fit model because of negligible star formation in low-mass
haloes due to radiative feedback. 

\item For low star-forming efficiencies of PopIII stars ($\sim 0.01$), the
reionization is actually driven by PopII stars and occurs rapidly and
shortly before $z=6$. In this case, the sensitivity required 
to observe the reionization sources 
(PopII stars within haloes of masses $< 10^8 M_{\odot}$)
at $z \gtrsim 6$ would be 33 AB magnitude.

\end{itemize}

The sensitivities required to observe the reionization sources
are expected to be achieved in future deep imaging surveys, particularly
the ones which take advantage of gravitational lensing magnification. 
For example, the 
NIR Wide Field Camera 3 
(WFC3), scheduled to be
installed on HST in near future, promises to achieve a
sensitivity limit of $m_{\rm AB} \sim 31$ in the 
$J_{110}$ and $H_{160}$ filters for a field of view of
$\sim 5$ arcmin$^2$ in about a few hundred hours
of observation time \cite{sle07}. According to our estimates, such 
lensed surveys should be able
to detect quite a few PopIII reionization sources 
(via Lyman-break dropout techniques) in the field of view. 
A much better prospect of detecting these sources would
be through the Ultra-Deep Imaging
Survey using the JWST which too plans to achieve a
sensitivity limit of $m_{\rm AB} \gtrsim 31$ over 100-200 hours
of observation time per filter \cite{gmc++06}.
Direct detection of these sources would put stringent constraints
on reionization history, and in addition can be used for understanding
physical processes like feedback.

Provided that the contamination problems due to bright atmospheric emission
lines, which could restrict visibility up to 50\% of the redshift range in the
$J$-band, could be maintained under control, a complementary approach for detecting 
the reionization sources would be through narrow band surveys for Ly$\alpha$ emitters. A rough
estimate of the Ly$\alpha$ luminosity from a halo of mass $M$
forming PopIII stars is given by
\be
L_{\alpha} = \epsilon_{*,{\rm III}} (1 - f_{\rm esc,III}) 
\f{\Omega_b}{\Omega_m} M ~ c_{{\rm Ly}\alpha} ~ q(H)
\e
where $c_{{\rm Ly}\alpha} = 1.04 \times 10^{11}$ ergs and 
$q(H)$ is the rate of hydrogen-ionizing photons per unit mass
of stars formed \cite{schaerer02}. 
The above relation does not take into account the attenuation
arising from neutral hydrogen around the source, hence the luminosities
could possibly be overestimated. Under the above assumptions, 
a $10^8 M_{\odot}$ halo would produce a luminosity of 
$\sim 10^{41}$ erg s$^{-1}$ in our best-fit model. Such luminosities
seem to be well within the reach of lensed Ly$\alpha$ surveys.
For example, \citeN{ser++07} have
detected $\sim 2$ sources having luminosities $\sim 10^{41.5}$ erg s$^{-1}$
at $z \sim 10$ within a area of 0.3 arcmin$^2$ using a Keck survey of gravitationally-lensed sources. There is a possibility that these
sources could be PopIII stars forming in haloes 
of masses $\gtrsim 10^8 M_{\odot}$ which do have $\lesssim 30$\% contribution 
to the ionizing power at $z \approx 10$. 
Future narrowband Ly$\alpha$ surveys 
like DAzLE\footnote{http://www.ast.cam.ac.uk/$\sim$optics/dazle/} thus would
be extremely important for direct detection of primary reionization sources
at high redshift.

Finally, it is worth mentioning that our models do not 
include (i) star formation in minihaloes where molecular cooling
is efficient, or (ii) the possibility of a top-heavy IMF
for PopIII stars. The predictions regarding the source counts
could vary considerably depending on the details of the above two 
processes, and hence future surveys present an excellent opportunity
to probe such effects.

\vspace{-0.7cm}
\section*{Acknowledgments}
We are indebted to R. Salvaterra and R. Schneider for useful discussions and for sharing their results.

\vspace{-0.7cm}
\bibliography{mnrasmnemonic,astropap-mod,reionization}
\bibliographystyle{mnras}

\end{document}